\documentclass[fleqn,usenatbib]{mnras}
\usepackage{newtxtext,newtxmath}
\usepackage[T1]{fontenc}
\usepackage{ae,aecompl}
\usepackage{graphicx}	
\usepackage{amsmath}
\usepackage{amssymb}
\usepackage{times}
\usepackage{subfigure} 

\title[$V_{\rm rms}$ Gradient in MaNGA]{SDSS-IV MaNGA: Stellar population correlates with stellar root-mean-square velocity $V_{\rm rms}$ gradients or total-density-profile slopes at fixed effective velocity dispersion $\sigma_{\rm e}$}
\author[S. Lu et al.]
{Shengdong Lu$^{1,2}$\thanks{E-mail: \url{lushengdong@nao.cas.cn}},
Michele Cappellari$^{3}$,
Shude Mao$^{4,1}$,
Junqiang Ge$^{1}$,
Ran Li$^{1,5}$
\\
$^{1}$National Astronomical Observatories, Chinese Academy of Sciences, 20A Datun Road, Chaoyang District, Beijing 100101, China\\
$^{2}$University of Chinese Academy of Sciences, Beijing 100049, China\\
$^{3}$Sub-department of Astrophysics, Department of Physics, University of Oxford, Denys Wilkinson Building, Keble Road, Oxford, OX1 3RH, UK\\
$^{4}$Department of Astronomy and Tsinghua Center for Astrophysics, Tsinghua University, Beijing 100084, China\\
$^{5}$School of Astronomy and Space Science, University of Chinese Academy of Science, Beijing 100049, China
}

\date{Accepted ***. Received ***; in original form ***}

\pubyear{2020}

\begin{document}
\label{firstpage}
\pagerange{\pageref{firstpage}--\pageref{lastpage}}
\maketitle
\begin{abstract}
Galaxy properties are known to correlate most tightly with the galaxy effective stellar velocity dispersion $\sigma_{\rm e}$. Here we look for {\em additional} trends at fixed $\sigma_{\rm e}$ using 1339 galaxies ($M_\ast \gtrsim 6\times10^9$ M$_\odot$) with different morphologies in the MaNGA (DR14) sample with integral-field spectroscopy data. We focus on the gradients ($\gamma_{\rm rms} \equiv \sigma(R_{\rm e}/4)/\sigma_{\rm e}$) of the stellar root-mean-square velocity ($V_{\rm rms} \equiv \sqrt{V^2 + \sigma^2}$), which we show traces the total mass density gradient $\gamma_{\rm tot}$ derived from dynamical models and, more weakly, the bulge fraction. We confirm that $\gamma_{\rm rms}$ increases with $\sigma_{\rm e}$, age and metallicity. We additionally find that these correlations still exist at fixed $\sigma_{\rm e}$, where galaxies with larger $\gamma_{\rm rms}$ are found to be older and more metal-rich. It means that mass density gradients contain information of the stellar population which is not fully accounted for by  $\sigma_{\rm e}$. This result puts an extra constraint on our understanding of galaxy quenching. We compare our results with galaxies in the IllustrisTNG hydrodynamical simulations and find that, at fixed $\sigma_{\rm e}$, similar trends exist with age, the bulge fraction, and the total mass density slope but, unlike observations, no correlation with metallicity can be detected in the simulations.
\end{abstract}


\section{Introduction}
\label{sec:introduction}
Galaxy properties have already been found to strongly correlate with their morphologies. One of the quantities that are used to describe the morphology of galaxies is the bulge fraction. However, the photometric bulge-disk decomposition is a complex process, which depends on the extraction details and suffers from degeneracies (\citealt{Cappellari_et_al.(2013b)}, hereafter C13b). Due to the degeneracies of projection, the intrinsic 3D stellar density cannot be uniquely recovered from the 2D images unless the galaxies are edge-on, even if we assume galaxies are axisymmetric \citep{Rybicki(1987),Gerhard_et_al.(1996),van_den_Bosch(1997),Magorrian_et_al.(1998)}. The fact that one cannot infer the intrinsic density of inclined galaxies implies that one cannot uniquely decompose their bulges and disks of inclined galaxies ($i<90^{\circ}$) unless we know the galaxies to be accurately described by \cite{Sersic(1968)} bulges and exponential disks \citep{Freeman(1970)}. \citet{Cappellari_et_al.(2011b)} used the \emph{kinematic morphology} (i.e. fast and slow rotators) to study the relation between galaxy type and the environment density (i.e. the $T-\Sigma$ relation), instead of using classic morphology (i.e. lenticulars and ellipticals). It is found that the $T-\Sigma$ relation is cleaner when the kinematic morphology is adopted. It implies that combining dynamical properties may be beneficial for analyzing the structure of galaxies. In C13b, $\sigma_{\rm e}$ (the velocity dispersion within the half-light radius $R_{\rm e}$) is shown to be related to the bulge fraction of galaxies (see fig. 5 of C13b).

\citet{Cappellari_et_al.(2006)} found that the stellar mass-to-light ratio $M_{\ast}/L$, which is related to the stellar population, tightly correlates with $\sigma_{\rm e}$ rather than mass or $R_{\rm e}$. Similarly, \citet{Graves_et_al.(2009)} used the Fundamental Plane to conclude that no stellar population property shows any dependence on $R_{\rm e}$ at fixed $\sigma_{\rm e}$. The primary role of $\sigma_{\rm e}$ in driving variations in both the stellar population indicators ($M_{\ast}/L$, colors, and line indices) and the molecular gas content was clearly demonstrated by C13b and with population models by \citet{McDermid_et_al.(2015)} using the high-quality data from the ATLAS$^{\rm 3D}$ survey. The latter ATLAS$^{\rm 3D}$ works showed that all population and gas content indicators closely follow lines of constant $\sigma_{\rm e}$ on the mass-size plane. It indicates that $\sigma_{\rm e}$, which traces the bulge fraction, is the main driver for the observed trends of the stellar population properties on the mass-size plane. This result is verified with a larger sample of galaxies with different morphologies from SAMI \citep[1319 galaxies]{Scott_et_al.(2017)} and MaNGA \citep[$\sim 2000$ galaxies]{Li_et_al.(2018)}.

Given that galaxy properties correlate with $\sigma_{\rm e}$ and that this in turn appears related to the galaxies bulge fraction, or the steepness of the total density profile, it is natural to ask whether other quantities contain extra information on the population that is not already contained in $\sigma_{\rm e}$ alone. More specifically, we want to see whether there is a residual correlation of stellar population properties with the root-mean-aquare velocity ($V_{\rm rms}\equiv \sqrt{V^2+\sigma^2}$, where $V$ and $\sigma$ are the mean line-of-sight velocity and dispersion) profile at fixed $\sigma_{\rm e}$. In C13b, the shape of $V_{\rm rms}$ map is proved to have the ability to trace the bulge fraction at even low inclinations (close to face-on, $i\sim 30^{\circ}$, see fig. 4 of C13b). The gradient of $V_{\rm rms}$ (defined there as $\sigma(R_{\rm e}/8)/\sigma_{\rm e}$, where $\sigma(R_{\rm e}/8)$ is the velocity dispersion within $R_{\rm e}/8$), which encodes key information of the shape of the $V_{\rm rms}$ map (see Fig.~\ref{fig:example} for some examples of the $V_{\rm rms}$ maps of MaNGA galaxies), is found to have similar distribution as stellar population properties on the mass-size plane, which implies the correlations between the gradient of $V_{\rm rms}$ and stellar populations. Notice that both $\sigma_{\rm e}$, $\sigma(R_{\rm e}/8)$, and $\sigma(R_{\rm e}/4)$ in later text (see Eq.~\ref{eq:eq2}) are all luminosity-weighted $V_{\rm rms}$, not purely dispersion $\sigma$ in given apertures (see Eq.~\ref{eq:eq3} for definition). However the ATLAS$^{\rm 3D}$ sample only contains 260 galaxies and does not allow one to asses whether a residual trends exists at fixed $\sigma_{\rm e}$, due to the small-numbers statistics (see \citealt{Cappellari(2016)} for a review). 

With the advancement of the largest Integral Field Unit (IFU) survey MaNGA \citep{Bundy_et_al.(2015)}, we are able to study the underlying relations between the kinematics and the stellar populations in detail with a sufficiently large number of galaxies with different morphologies. The goals of this paper are: (\textbf{i}) to investigate the correlations between the gradient of $V_{\rm rms}$ and galaxy properties (i.e. age, metallicity, the bulge fraction, and the steepness of total mass profile); (\textbf{ii}) to examine whether the correlations still exist at fixed $\sigma_{\rm e}$ which has not been addressed before; (\textbf{iii}) to compare with the state-of-the-art hydrodynamical simulations, the IllustrisTNG simulations \citep{Marinacci_et_al.(2018),Naiman_et_al.(2018),Nelson_et_al.(2018),Pillepich_et_al.(2018),Springel_et_al.(2018)}. 

This paper is organized as follows. In Section~\ref{sec:data}, we describe our sample selection in MaNGA (Section~\ref{sec:manga}) and IllustrisTNG simulations (Section~\ref{sec:tng}). The general property of the $V_{\rm rms}$ gradient is presented in Section~\ref{sec:result1}. Section~\ref{sec:result2} is devoted to presenting the correlations between the gradient of $V_{\rm rms}$ and galaxy properties. In Section~\ref{sec:result3}, we investigate the relations between the gradient of $V_{\rm rms}$ and galaxy properties at fixed $\sigma_{\rm e}$. Finally, we summarize our findings in Section~\ref{sec:conclusion}.

\section{Data}
\label{sec:data}
\subsection{MaNGA galaxies}
\label{sec:manga}
The galaxies in this study are from the MaNGA sample released by SDSS DR14 \citep{Abolfathi_et_al.(2018)}, which includes 2778 galaxies with different morphologies. We note here that we do not use the latest data release of MaNGA because we want our results to rely on published data (i.e. stellar population and structural properties) from other papers. The kinematical data are extracted from the IFU spectra using the MaNGA data analysis pipeline (DAP; \citealt{Westfall_et_al.(2019)}) by fitting absorption lines, making use of the {\sc ppxf} software \citep{Cappellari_and_Emsellem(2004),Cappellari(2017)} with a subset of the {\sc MILES} \citep{Sanchez-Blazquez_et_al.(2006),Falcon-Barroso_et_al.(2011)} stellar library, {\sc MILES-THIN}. Before fitting, the spectra are Voronoi binned \citep{Cappellari_and_Copin(2003)} to $\mathrm{S/N}=10$. Readers are referred to the following papers for more details on the MaNGA instrumentation \citep{Drory_et_al.(2015)}, observing strategy \citep{Law_et_al.(2015)}, spectrophotometric calibration \citep{Smee_et_al.(2013),Yan_et_al.(2016a)}, and survey execution and initial data quality \citep{Yan_et_al.(2016b)}. 

We exclude the galaxies which are merging or have low data quality (with fewer than 100 Voronoi bins with $\mathrm{S/N}$ greater than 10) from our sample. Besides, we exclude galaxies with low stellar mass ($M_{\ast}<6\times 10^9\,\mathrm{M_{\odot}}$) to achieve a comparable minimum stellar mass as in C13b. The stellar mass of MaNGA galaxies is derived from \citet[table 1]{Salim_et_al.(2016)}, in which the state-of-the-art spectral energy distribution (SED) fitting of UV and optical fluxes is adopted. After excluding the galaxies described above, we have 1520 galaxies with different morphologies. We derive the stellar population properties (i.e. age and metallicity) from \citet[see their online table A1]{Li_et_al.(2018)}, which are calculated as the luminosity-weighted values within an elliptical aperture of area $A=\pi R_{\rm e}^2$. We use the mass-weighted total density slope ($\gamma_{\rm tot}$) from \citet[a subset\footnote{Readers need to cite \citet{Li_et_al.(2019)} if they want to use the total density slope in this work.} of the values for our galaxies is included in Table~\ref{table:table1}]{Li_et_al.(2019)} to describe the total mass profiles of galaxies (see \citealt[eq. 2]{Li_et_al.(2019)} for definition). The bulge-to-total luminosity ratio ($\mathrm{B/T}$) is from \citet[table 1]{Simard_et_al.(2011)}. After cross-matching these four catalogs, we have 1339 galaxies with available galaxy properties in our final sample. The size parameters (i.e. the half-light radius $R_{\rm e}$ and the major axis of the half-light isophote $R_{\rm e}^{\rm maj}$) used in this work are calculated from the Multi-Gaussian Expansion (MGE) models  \citep{Emsellem_et_al.(1994)} with the fitting algorithm and {\sc python} software\footnote{The software is available from \url{http://www-astro.physics.ox.ac.uk/~mxc/software}} by \citet{Cappellari(2002)} and are scaled by a factor of 1.35 (see fig. 7 of \citealt{Cappellari_et_al.(2013a)}).

\subsection{Galaxies in hydrodynamical simulations}
\label{sec:tng}
For comparison, we also select a sample of simulated galaxies with the same stellar mass range ($M_{\ast}>6\times 10^9\,\mathrm{M_{\odot}}$) at $z=0$ from the state-of-the-art magneto-hydrodynamic cosmological galaxy formation simulations, the IllustrisTNG simulations\footnote{\url{http://www.tng-project.org}} (TNG hereafter; \citealt{Marinacci_et_al.(2018),Naiman_et_al.(2018),Nelson_et_al.(2018),Pillepich_et_al.(2018),Springel_et_al.(2018)}). In this work, we use its full-physics version with a cubic box of $110.7\,\mathrm{kpc}$ side length (TNG100) which has been made publicly available \citep{Nelson_et_al.(2019)}. The mass resolutions of the TNG100-full physics version for baryonic and dark matter are $m_{\rm baryon}=1.4\times10^6\,{\rm M_{\odot}}$ and $m_{\rm DM}=7.5\times10^6\,{\rm M_{\odot}}$, with a gravitational softening length of $\epsilon = 0.74\,\mathrm{kpc}$. Gas cells are resolved in a fully adaptive manner with a minimum softening length of $0.19$ comoving $\mathrm{kpc}$. All selected galaxies are central galaxies whose host dark matter subhaloes are identified by the {\sc subfind} algorithm \citep{Springel_et_al.(2001),Dolag_et_al.(2009)}. Thus, we have 5105 simulated galaxies in our sample.

To derive stellar population properties of simulated galaxies, we first project them along the X-axis of the simulation box to produce mock images in the SDSS $r$-band~\citep{Stoughton_et_al.(2002)} filter. We then use the MGE method to obtain their size parameters as for real galaxies. The age and metallicity of TNG galaxies are calculated as luminosity-weighted $\log\,\mathrm{Age}$ and $\mathrm{[Z/H]}$ within an elliptical aperture of area $A=\pi R_{\rm e}^2$ using the equation below:
\begin{equation}
\label{eq:eq1}
\langle x\rangle = \frac{\Sigma_{k}L_{k}x_{k}}{\Sigma_{k}L_{k}},
\end{equation}
where $x_{k}$ is $\log\,\mathrm{Age}$ (or $\mathrm{[Z/H]}$) of the $k$-th particle within this elliptical aperture, and $L_{k}$ is the SDSS $r$-band luminosity of the $k$-th particle. The bulge-to-total ratio of TNG galaxies is from \citet{Xu_et_al.(2019)}. We refer the readers to \citet{Xu_et_al.(2017)} for detailed descriptions of galaxy property extraction. The total mass density slopes of TNG galaxies are from \citet{Li_et_al.(2019)}. All these related properties of TNG galaxies can be derived from the journal website and readers are required to cite \citet{Xu_et_al.(2019)} if they need to use the bulge fraction in this work.

We note here that we want to make an impeccable comparison with the IllustrisTNG simulations, and as a result, we exclude the galaxies with low stellar mass ($M_{\ast}<6\times 10^9\,\mathrm{M_{\odot}}$) in MaNGA. This mass limit roughly matches the mass cutoff of the IllustrisTNG simulations ($\sim 5\times 10^9\,\mathrm{M_{\odot}}$, below which the galaxies may not be sufficiently resolved). We have also examined our results with all qualified galaxies in MaNGA regardless of their stellar masses and find our results remain unchanged.

\section{Results}
\label{sec:results}
In this section, we first present the distribution of the gradient of $V_{\rm rms}$ in Section~\ref{sec:result1}. Then we show the relations between the gradient of $V_{\rm rms}$ and galaxy properties (i.e. $\mathrm{B/T}$, $\gamma_{\rm tot}$, $\log\mathrm{Age}$, and $\mathrm{[Z/H]}$) in Section~\ref{sec:result2}. Section~\ref{sec:result3} is devoted to presenting the relation between the gradient of $V_{\rm rms}$ and galaxy properties at fixed $\sigma_{\rm e}$.

\subsection{The gradient of $V_{\rm rms}$}
\label{sec:result1}
\begin{figure*}
\subfigure[1-211082]{\includegraphics[width=0.65\columnwidth]{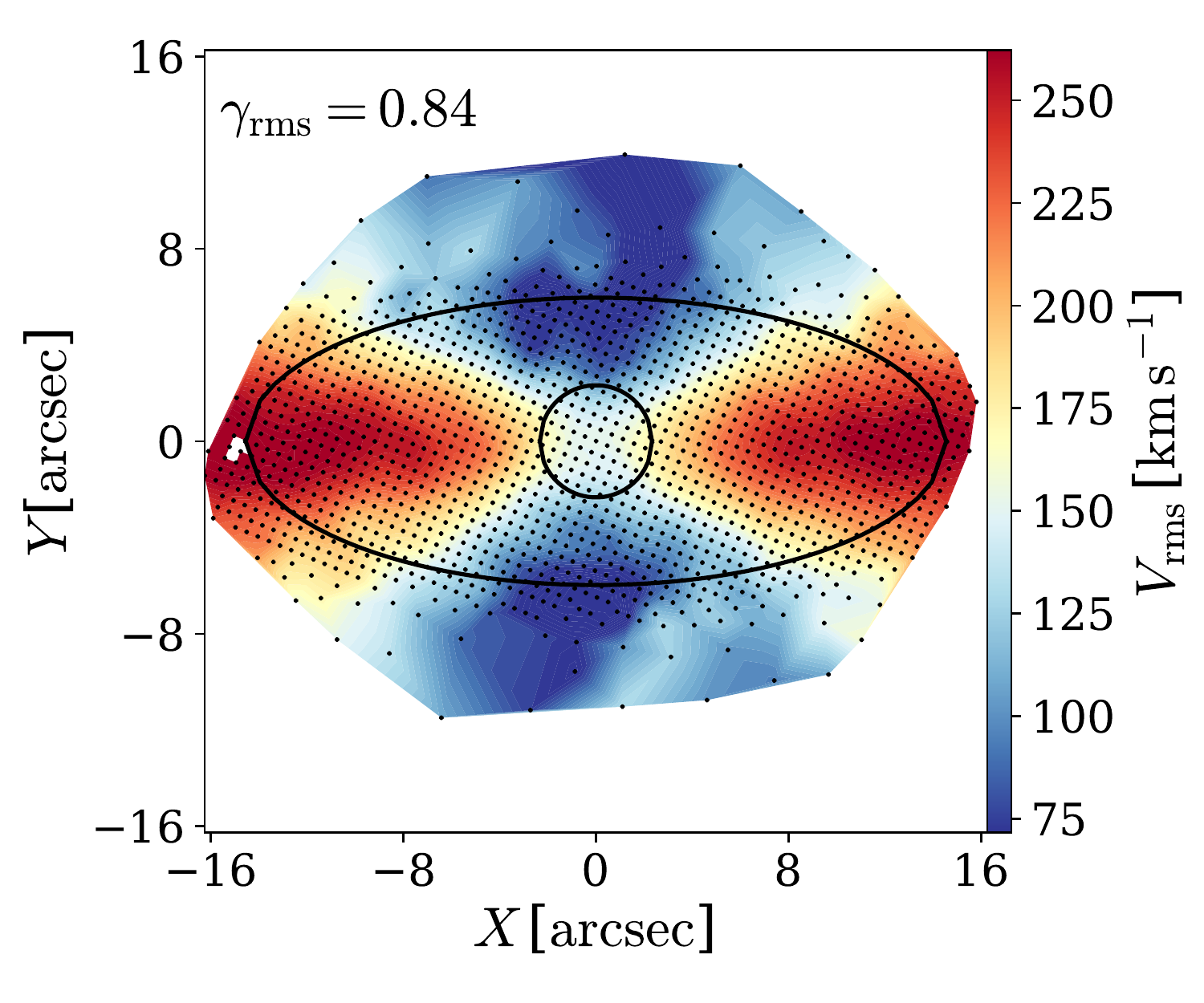}}
\subfigure[1-114145]{\includegraphics[width=0.65\columnwidth]{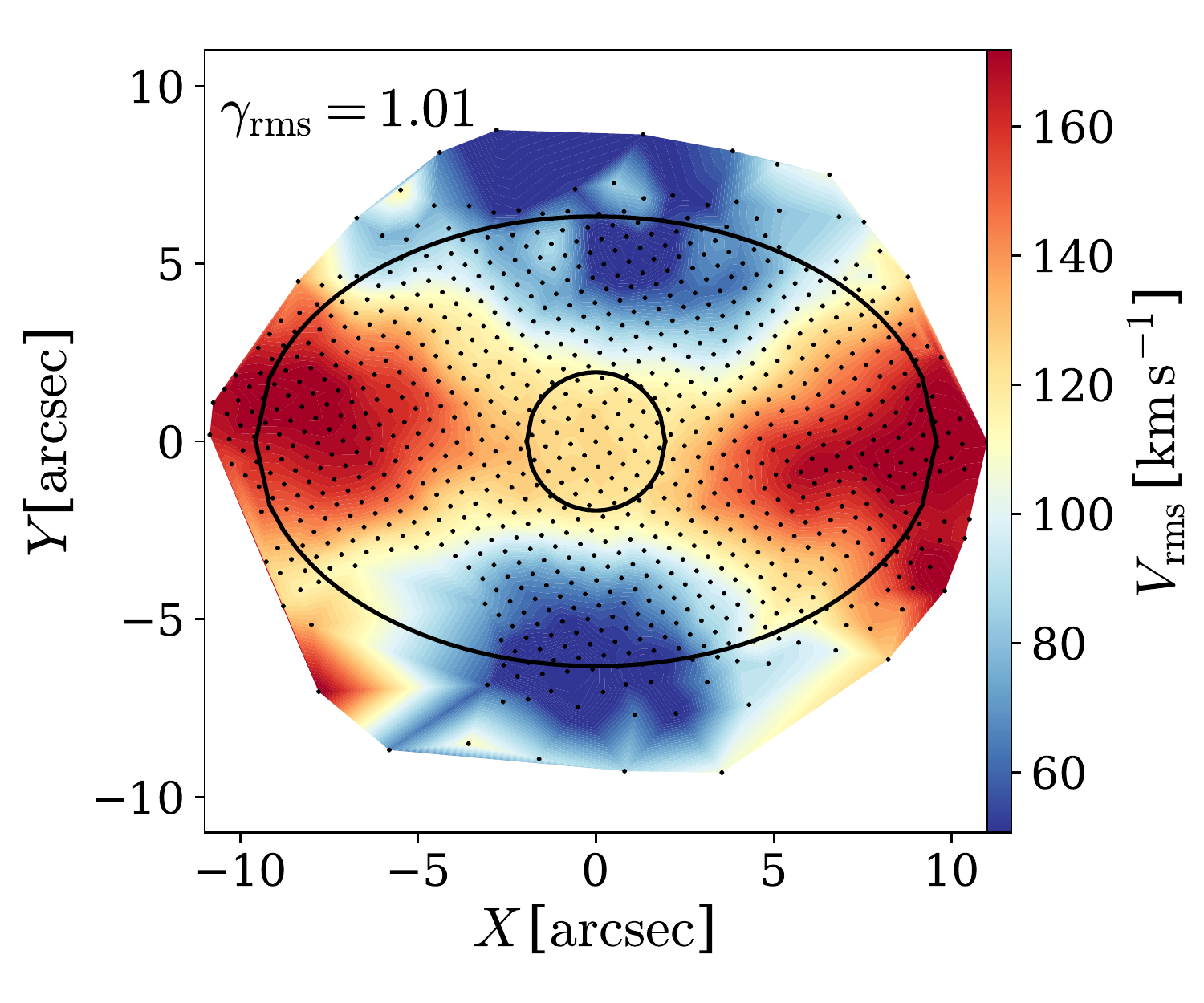}}
\subfigure[1-48208]{\includegraphics[width=0.65\columnwidth]{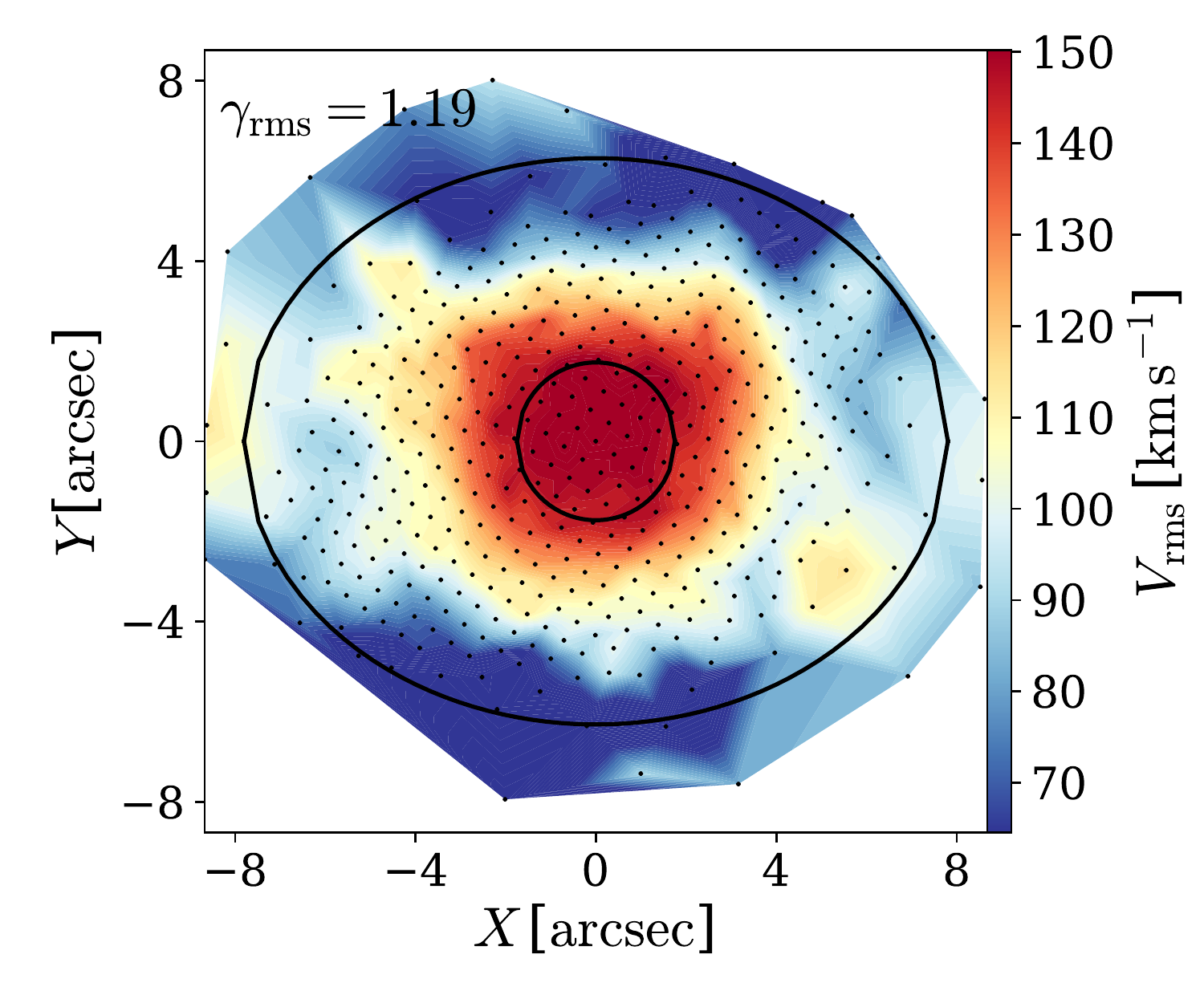}}
\caption{Three examples (MaNGA ID: 1-211082, 1-114145, 1-48208) of the $V_{\rm rms}$ ($\equiv\sqrt{V^2+\sigma^2}$) maps of MaNGA galaxies with different $\gamma_{\rm rms}$. In each panel, the outer ellipse encloses an area $A=\pi R_{\rm e}^2$, while the inner circular has radius $R_{\rm e}/4$.}
\label{fig:example}
\end{figure*}

Similarly to C13b, we define the gradient of $V_{\rm rms}$ as:
\begin{equation}
\label{eq:eq2}
    \gamma_{\rm rms} \equiv \frac{\sigma(R_{\rm e}/4)}{\sigma_{\rm e}}.
\end{equation}
$\sigma_{\rm e}$ is the luminosity-weighted $V_{\rm rms}$ within an elliptical aperture of area $A=\pi R_{\rm e}^2$ and is calculated as:
\begin{equation}
\label{eq:eq3}
\sigma_{\rm e} = \sqrt{\frac{\Sigma_{k}F_{k}V_{\mathrm{rms},k}^2}{\Sigma_{k}F_{k}}} = \sqrt{\frac{\Sigma_{k}F_{k}(V_{k}^2+\sigma_{k}^2)}{\Sigma_{k}F_{k}}},
\end{equation}
where $V_{k}$ and $\sigma_{k}$ are the mean velocity and dispersion in the $k$-th IFU spaxel, and $F_{k}$ is the flux in the $k$-th spaxel. The sum is within the elliptical aperture described above. $\sigma(R_{\rm e}/4)$ is calculated in the same way but the sum is within a {\em circular} aperture with radius of $R_{\rm e}/4$ (see Fig.~\ref{fig:example} for examples). The $\sigma(R_{\rm e}/4)$ and $\sigma_{\rm e}$ so defined agree closely with the velocity dispersions measured from a single fit to the spectrum inside the same apertures in C13b \citep{Li_et_al.(2018)}. If a galaxy has $\gamma_{\rm rms}>1$, that means there is a decrease of $V_{\rm rms}$ from the inner to the outer region of this galaxy. We note here that due to the limitation of spatial resolution in MaNGA, we use a larger ($R_{\rm e}/4$ instead of $R_{\rm e}/8$) aperture for the central velocity dispersion compared to that in C13b. We show some examples of the $V_{\rm rms}$ maps of MaNGA galaxies in Fig.~\ref{fig:example}, from which we can see an obvious change of the map shape from low $\gamma_{\rm rms}$ to high $\gamma_{\rm rms}$ (see Section~\ref{sec:result2} for more detailed discussion). Part of the quantitative results are listed in Table~\ref{table:table1}. For simulated galaxies, $\langle v_{\rm los}^2\rangle^{1/2}$ is used to approximate the $V_{\rm rms}$. 

Due to the effect of seeing, we have tried to only use galaxies with $R_{\rm e}/4>1.5^{\prime\prime}$, where $1.5^{\prime\prime}$ is the typical value of seeing in MaNGA (see \citealt{Bundy_et_al.(2015)} for more information of technical details of MaNGA), which constitute about half of our current sample (red circles with green dots in the center in Fig.~\ref{fig:gamma_sigma} and Fig.~\ref{fig:gamma_galaxy}). Our main results are found to be unchanged with only large galaxies. Besides, we have also tried to use the linear slopes of $\log\,V_{\rm rms}$ profiles ($\log\,V_{\rm rms}$ versus $\log\,R$) to describe the gradient of $V_{\rm rms}$. We first divide the galaxies into several elliptical annuli from $R_{\rm e}/8$ to $R_{\rm e}$, with the global ellipticity measured around $1R_{\rm e}$. Then we calculate the median value of $\log\,V_{\rm rms}$ in each annulus and perform a linear fit to get the linear slopes. We find that our main results remain unchanged when the new gradient is adopted.

\begin{figure}
\centering
\includegraphics[width=\columnwidth]{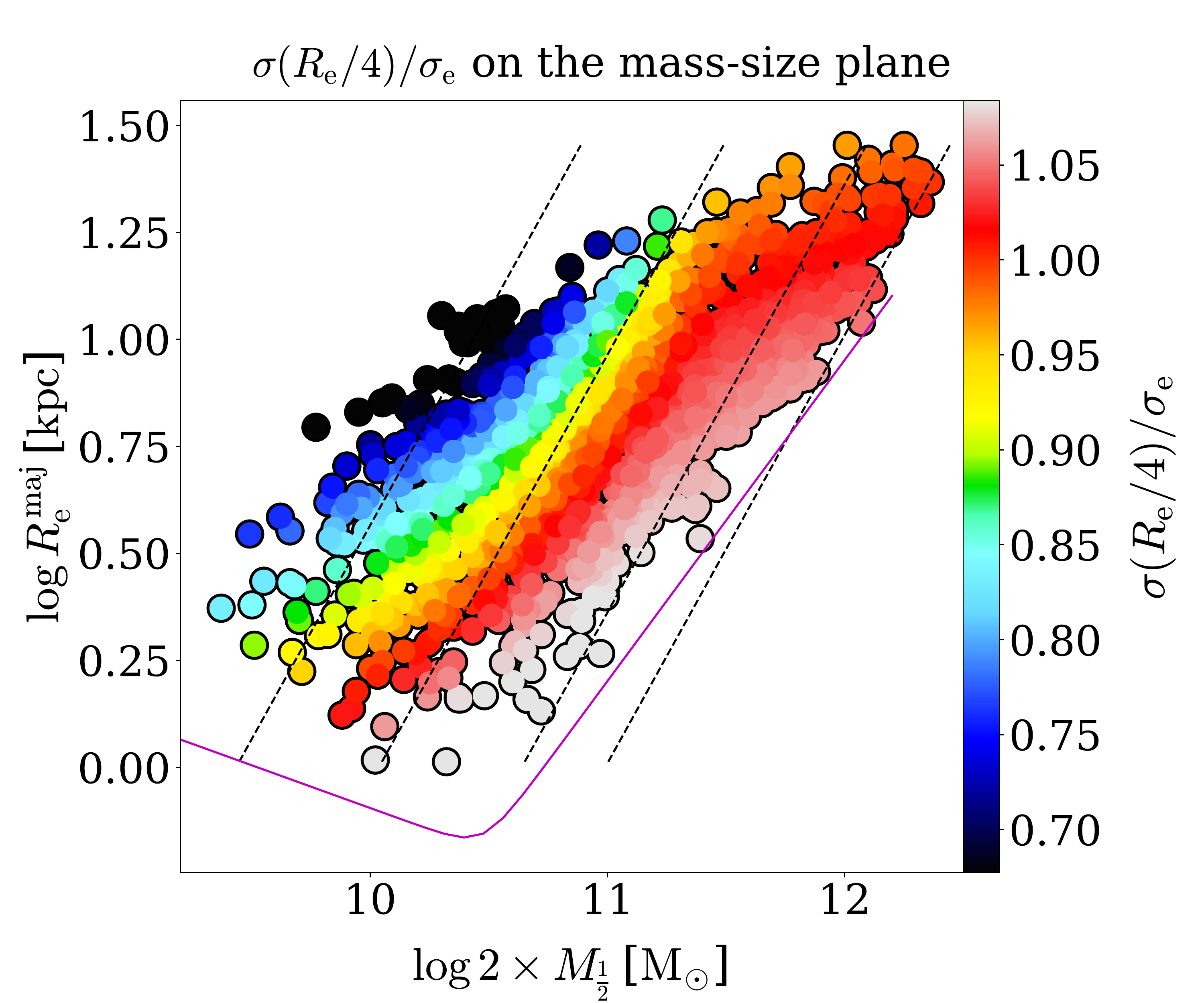}
\caption{The distribution of MaNGA galaxies on the mass-size plane ($R_{\rm e}^{\rm maj}$ vs. $M_{1/2}$), color-coded by the $V_{\rm rms}$ gradient $\gamma_{\rm rms}$ ($\equiv \sigma(R_{\rm e}/4)/\sigma_{\rm e}$). The black dashed lines are the lines of constant $\sigma_{\rm e}$: 50, 100, 200, and 300 $\mathrm{km\,s^{-1}}$. The magenta curve is the zone of exclusion defined in C13b.} 
\label{fig:ms}
\end{figure}

\begin{figure}
\centering
\includegraphics[width=\columnwidth]{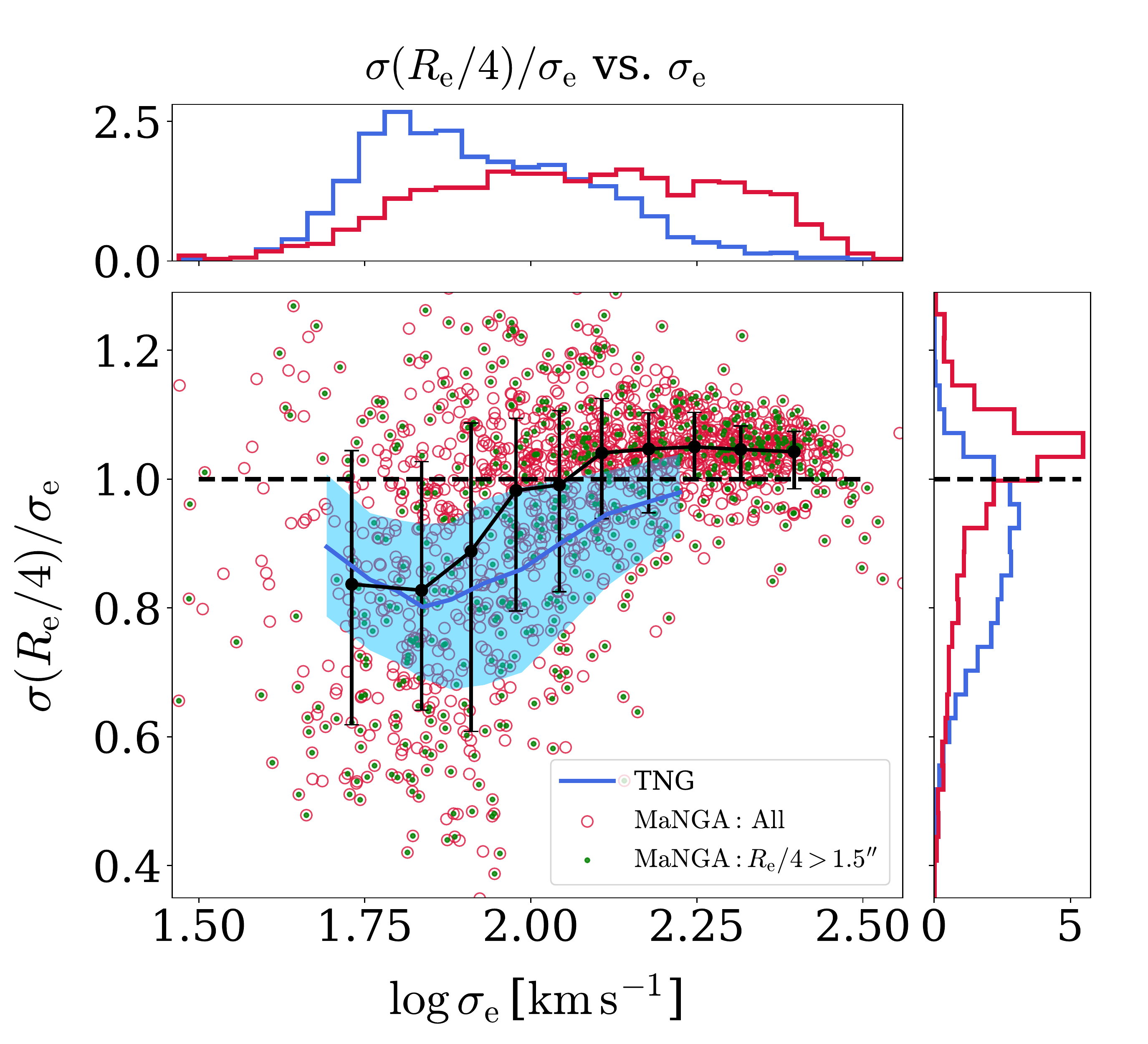}
\caption{The relation between the $V_{\rm rms}$ gradient $\gamma_{\rm rms}$ and $\sigma_{\rm e}$. The red circles are MaNGA galaxies and the circles with green dots in the center are the galaxies with $R_{\rm e}/4>1.5^{\prime\prime}$, where $1.5^{\prime\prime}$ is the typical value of seeing in MaNGA. The black dots represent the median values of $\gamma_{\rm rms}$ in each $V_{\rm rms}$ bin, with error bars indicating the 1$\sigma$ range. TNG galaxies are shown by the median profile (blue lines) with a shaded region indicating the 1$\sigma$ range. The black dashed line represents $\gamma_{\rm rms}=1$, below which an increase of $V_{\rm rms}$ exists in galaxies. The histograms are the distributions of $\gamma_{\rm rms}$ and $\log\,\sigma_{\rm e}$ for MaNGA (red) and TNG (blue) galaxies.} 
\label{fig:gamma_sigma}
\end{figure}

We show the distribution of $\gamma_{\rm rms}$ ($\equiv \sigma(R_{\rm e}/4)/\sigma_{\rm e}$) on the $R_{\rm e}^{\rm maj}$-$ M_{1/2}$ plane (the `mass-size plane') in Fig.~\ref{fig:ms}, where $M_{1/2}$ is the enclosed total mass within a spherical aperture of the three-dimensional half-light radius. The $M_{1/2}$ here is from \citet{Li_et_al.(2018)} which is derived using the Jeans anisotropic model (JAM) \citep{Cappellari(2008)}. Before plotting, we make use of the {\sc python} implementation\footnote{The software is available from \url{https://pypi.org/project/loess/}} (see details in C13b) of the two-dimensional Locally Weighted Regression (LOESS) \citep{Cleveland_and_Devlin(1988)} method to obtain smoothed distribution of $\gamma_{\rm rms}$. As can be seen, the gradient of $V_{\rm rms}$ varies systematically roughly along the $\sigma_{\rm e}$ direction, which is consistent with the result in C13b (the top panel of their fig. 6) in which $\sigma(R_{\rm e}/8)/\sigma_{\rm e}$ is used. We note here that due to the fact that we use a different definition of the gradient of $V_{\rm rms}$ from C13b and our samples include late-type galaxies which typically have lower $\gamma_{\rm rms}$, the smoothed $\gamma_{\rm rms}$ values obtained with the LOESS method in this figure are somehow lower than those in C13b.

To better illustrate the correlation between $\gamma_{\rm rms}$ and $\sigma_{\rm e}$, we present Fig.~\ref{fig:gamma_sigma}. As shown in the figure, $\gamma_{\rm rms}$ of MaNGA galaxies increases with $\sigma_{\rm e}$ from $\gamma_{\rm rms}<1$ (at low $\sigma_{\rm e}$) to $\gamma_{\rm rms}>1$ (at high $\sigma_{\rm e}$). The transition point $\gamma_{\rm rms}=1$, which corresponds to a flat $V_{\rm rms}$ profile, is at $\log \sigma_{\rm e}\approx 2.1$. For $\log\,\sigma_{\rm e}\gtrsim 2.1$, the trend becomes flatter and $\gamma_{\rm rms}$ stays at $\sim 1.05$, indicating a mild decrease of $V_{\rm rms}$ from the central region of galaxies to the outer part. As demonstrated in C13b (fig. 5), galaxies with $\log \sigma_{\rm e}\approx 2.0$ have $V_{\rm rms}$ maps in butterfly-like shapes with small bulges in the center, after which the peak of $V_{\rm rms}$ in the center gets stronger, indicating a rising bulge fraction. It implies that the relation between the gradient of $V_{\rm rms}$ and the bulge fraction saturates when this is larger than a threshold. In the idealized case in which the bulge and disk have fixed parametrization (e.g. exponential and S{\'e}rsic profiles), this can be qualitatively understood as due to the fact that, when the bulge dominates, the $V_{\rm rms}$ gradient must converge to that of the bulge alone. The most massive galaxies ($M_{\ast}>2\times10^{11}\,\mathrm{M_{\odot}}$), which also have large $\sigma_{\rm e}$, tend to be slow rotators without disks. For those galaxies, $V\sim0$ and the $V_{\rm rms}$ coincides with the $\sigma$, which has a characteristic decreasing radial profile with radius. This is illustrated in Fig.~\ref{fig:gamma_galaxy}, in which $\gamma_{\rm rms}$ increases with $\mathrm{B/T}$ and levels off when $\mathrm{B/T}\gtrsim 0.5$ (see Section~\ref{sec:result2} for a detailed description). This trend in $\gamma_{\rm rms}$ with $\sigma_{\rm e}$ is similar with the trend in total density slope $\gamma_{\rm tot}$ found by \citet{Poci_et_al.(2017)} and \citet{Li_et_al.(2019)}, which also shows a near constant trend above $\log \sigma_{\rm e}\approx 2.1$ and a break, and an increase in $\gamma_{\rm tot}$, below that transition value (see fig.1 of \citealt{Li_et_al.(2019)}). The key difference is that $\gamma_{\rm rms}$ is a purely empirical quantity, which does not involve dynamical models. As a comparison, however, TNG galaxies show a decreasing trend at low $\sigma_{\rm e}$ end and do not show a flat tail at high $\sigma_{\rm e}$ end. Besides, most ($\sim 86\%$) TNG galaxies have $\gamma_{\rm rms}<1$, relative to $44\%$ in MaNGA, and $\sigma_{\rm e}$ of TNG galaxies is typically lower than MaNGA galaxies. This may be due to the overly-strong AGN feedback (typically the isotropic black hole kinetic winds in the AGN quiescent phase) in IllustrisTNG for puffing up the galaxies \citep{Wang_Y._et_al.(2019),Wang_Y._et_al.(2020),Lu_et_al.(2020)}.

\begin{table}
\setlength{\tabcolsep}{1.3mm}
\begin{tabular}{lcccccc}
\hline
\hline
MaGNA ID & $\log\,R_{\rm e}$ & $\log\,R_{\rm e}^{\rm maj}$ & $\log\,\sigma(R_{\rm e}/4)$ & $\log\,\sigma_{\rm e}$ & $\gamma_{\rm rms}$ & $\gamma_{\rm tot}$\\
 & [$\mathrm{arcsec}$] & [$\mathrm{arcsec}$] & [$\mathrm{km\,s^{-1}}$] & [$\mathrm{km\,s^{-1}}$] & & \\
\hline
1-24295 & 0.85 & 0.85 & 2.27 & 2.16 & 1.05 & 1.95\\
1-260743 & 0.78 & 0.79 & 2.44 & 2.42 & 1.01 & 2.06\\
1-115128 & 0.90 & 1.07 & 1.83 & 1.97 & 0.93 & 1.83\\
1-258820 & 1.11 & 1.25 & 1.09 & 1.87 & 0.59 & 1.14\\
1-251662 & 0.76 & 0.78 & 2.24 & 1.98 & 1.13 & 1.78\\
1-637825 & 1.05 & 1.14 & 1.53 & 1.92 & 0.80 & 1.62\\
1-23877 & 1.02 & 1.31 & 1.70 & 2.09 & 0.82 & 1.60\\
1-109521 & 0.85 & 0.91 & 2.32 & 2.28 & 1.02 & 2.02\\
1-321967 & 0.39 & 0.45 & 1.88 & 1.88 & 1.00 & 1.74\\
1-147685 & 0.88 & 1.28 & 0.82 & 1.95 & 0.42 & 0.73\\
1-547185 & 0.93 & 0.99 & 1.76 & 1.83 & 0.96 & 1.84\\
1-38550 & 1.15 & 1.21 & 1.82 & 1.84 & 0.99 & 1.53\\
1-351790 & 0.62 & 0.73 & 1.56 & 1.66 & 0.94 & 1.50\\
1-258315 & 0.85 & 0.89 & 1.75 & 1.87 & 0.94 & 1.71\\
1-95770 & 0.85 & 0.95 & 1.64 & 1.87 & 0.88 & 0.98\\
1-22347 & 0.46 & 0.49 & 2.00 & 1.98 & 1.01 & 1.89\\
1-135091 & 0.87 & 0.94 & 2.28 & 2.22 & 1.02 & 2.05\\
1-260541 & 0.89 & 0.90 & 2.63 & 2.39 & 1.10 & 2.10\\
1-93908 & 0.85 & 0.91 & 2.58 & 2.33 & 1.11 & 2.35\\
1-275354 & 0.93 & 1.04 & 1.56 & 1.81 & 0.86 & 1.57\\
1-339116 & 0.67 & 0.81 & 1.34 & 1.81 & 0.74 & 1.30\\
1-167392 & 0.51 & 0.64 & 1.25 & 1.79 & 0.70 & 0.49\\
1-174036 & 0.78 & 0.82 & 2.27 & 2.37 & 0.95 & 1.80\\
1-55552 & 0.76 & 0.81 & 2.44 & 2.27 & 1.07 & 2.18\\
1-24476 & 0.96 & 1.05 & 2.19 & 2.08 & 1.05 & 2.01\\
1-545674 & 1.27 & 1.43 & 1.38 & 1.75 & 0.79 & 0.64\\
1-251278 & 0.66 & 0.70 & 2.54 & 2.35 & 1.08 & 2.20\\
1-594505 & 0.93 & 0.93 & 2.40 & 2.23 & 1.08 & 2.37\\
1-167334 & 0.60 & 0.68 & 2.06 & 2.00 & 1.03 & 2.39\\
1-217557 & 0.53 & 0.58 & 2.22 & 2.08 & 1.07 & 2.52\\
1-245908 & 0.57 & 0.68 & 1.99 & 1.95 & 1.02 & 1.96\\
1-44047 & 0.32 & 0.42 & 1.98 & 2.11 & 0.94 & 1.80\\
1-210667 & 1.12 & 1.25 & 1.64 & 1.76 & 0.93 & 1.28\\
1-73638 & 0.87 & 1.04 & 2.06 & 2.28 & 0.90 & 1.88\\
1-256465 & 0.61 & 0.66 & 2.38 & 2.28 & 1.04 & 2.42\\
1-114171 & 0.73 & 0.79 & 2.60 & 2.45 & 1.06 & 2.26\\
1-210961 & 0.58 & 0.86 & 1.89 & 2.03 & 0.93 & 2.16\\
1-217650 & 0.82 & 0.85 & 1.92 & 1.89 & 1.02 & 1.94\\
1-166739 & 0.83 & 0.88 & 2.40 & 2.32 & 1.04 & 2.26\\
1-164007 & 0.47 & 0.60 & 2.05 & 2.04 & 1.01 & 1.90\\
1-245940 & 0.91 & 1.03 & 2.36 & 2.52 & 0.93 & 2.03\\
1-274506 & 0.86 & 0.93 & 0.84 & 1.65 & 0.51 & 0.50\\
1-156011 & 0.75 & 0.75 & 2.57 & 2.09 & 1.23 & 2.31\\
1-210116 & 1.18 & 1.23 & 1.93 & 1.94 & 1.00 & 1.72\\
1-180208 & 0.55 & 0.67 & 2.44 & 2.27 & 1.07 & 2.90\\
12-84674 & 0.99 & 1.00 & 1.77 & 1.72 & 1.03 & 1.41\\
1-96075 & 1.05 & 1.11 & 1.82 & 2.07 & 0.88 & 1.61\\
1-23786 & 0.84 & 0.98 & 1.06 & 1.81 & 0.59 & 0.87\\
1-36977 & 0.78 & 0.81 & 2.20 & 2.12 & 1.04 & 2.00\\
1-457009 & 0.87 & 0.91 & 1.91 & 1.79 & 1.07 & 1.44\\
1-43272 & 0.47 & 0.53 & 1.57 & 1.81 & 0.87 & 1.22\\
12-84627 & 0.82 & 1.01 & 1.51 & 2.05 & 0.74 & 1.76\\
1-29726 & 0.68 & 0.77 & 2.55 & 2.44 & 1.04 & 2.28\\
1-38398 & 1.00 & 1.30 & 1.53 & 1.95 & 0.79 & 1.48\\
1-492524 & 0.79 & 1.03 & 1.69 & 1.83 & 0.92 & 1.67\\
1-273861 & 0.41 & 0.42 & 2.20 & 2.11 & 1.04 & 2.25\\
1-384548 & 0.43 & 0.56 & 2.33 & 2.22 & 1.05 & 2.56\\
\hline
\end{tabular}
\caption{MaNGA ID, half-light radius $\log\,R_{\rm e}$, major axis of the half-light isophote $\log\,R_{\rm e}^{\rm maj}$, luminosity-weighted velocity dispersion within two apertures (a circular aperture with radius $R_{\rm e}/4$ and an elliptical aperture of area $A=\pi R_{\rm e}^2$), the gradient of $V_{\rm rms}$ ($\gamma_{\rm rms} \equiv \sigma(R_{\rm e}/4)/\sigma_{\rm e}$), and the total density slope $\gamma_{\rm tot}$ (from \citealt{Li_et_al.(2019)}) for part of our sample. Please see the complete table on the journal website.}
\vspace{2mm}
\label{table:table1}
\end{table}

\subsection{$\gamma_{\rm rms}$ vs. galaxy properties}
\label{sec:result2}
\begin{figure*}
\centering
\includegraphics[width=2.1\columnwidth]{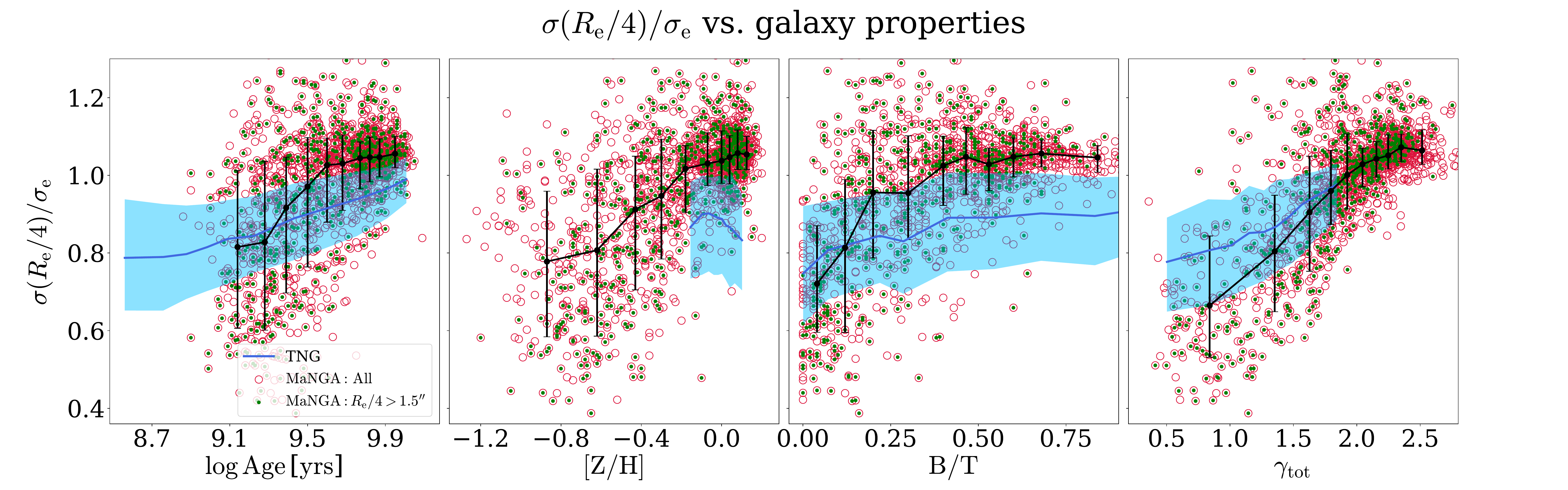}
\caption{The relation of $\gamma_{\rm rms}$ vs. $\log \mathrm{Age}$, $\mathrm{[Z/H]}$, $\mathrm{B/T}$, and the total density slope $\gamma_{\rm tot}$ (from left to right). The bulge fraction of MaNGA galaxies is derived from \citet[table 1]{Simard_et_al.(2011)}. The symbols are the same as Fig.~\ref{fig:gamma_sigma}.} 
\label{fig:gamma_galaxy}
\end{figure*}

To investigate the relation between the $V_{\rm rms}$ gradient and galaxy properties (i.e. the stellar age, metallicity, the total density slope, and the bulge fraction), we present Fig.~\ref{fig:gamma_galaxy}. As can be seen, $\gamma_{\rm rms}$ of MaNGA galaxies increases with $\log\mathrm{Age}$ and $\mathrm{[Z/H]}$, indicating that older and more metal-rich galaxies are more likely to have flatter or even declining $V_{\rm rms}$ profiles. For TNG galaxies, $\gamma_{\rm rms}$ shows a shallower correlation with $\log\mathrm{Age}$, while it does not show an obvious correlation with $\mathrm{[Z/H]}$. We note here that TNG lacks galaxies with low metallicity which results in the narrower range of metallicity.

Regarding the galaxy structural parameters, one can note that $\gamma_{\rm rms}$ ($\equiv \sigma(R_{\rm e}/4)/\sigma_{\rm e}$) increases with bulge fraction until $\mathrm{B/T}\sim 0.5$, but flattens for larger $\mathrm{B/T}$ values. This implies that either $\gamma_{\rm rms}$ is not a good indicator of bulge fraction for nearly pure spheroids (see also Section~\ref{sec:result1}), or that extreme bulge fractions are more difficult to measure from photometry. For the total density slope $\gamma_{\rm tot}$, which is defined as \citep[eq. 1]{Dutton_and_Treu(2014)}:
\begin{equation}
\gamma_{\rm tot} \equiv-\frac{1}{M\left(R_{\rm e}\right)} \int_{0}^{R_{\rm e}} 4 \pi r^{2} \rho(r) \frac{d \log \rho}{d \log r} d r=3-\frac{4 \pi R_{\rm e}^{3} \rho\left(R_{\rm e}\right)}{M\left(R_{\rm e}\right)},
\end{equation}
where $\rho(r)$ is total mass density of the galaxy and $M(R)$ is the total mass enclosed in a sphere with radius $R$ \citep{Li_et_al.(2019)}, the gradient of $V_{\rm rms}$ rises across the whole range of $\gamma_{\rm tot}$. It means that the steeper the mass profile, the more likely the $V_{\rm rms}$ profile to be flat or even declining. $\gamma_{\rm rms}$ of galaxies which are isothermal ($\gamma_{\rm tot}=2$) is $\sim 1$, indicating that flat $V_{\rm rms}$ profiles are exsitent in these galaxies. A purely isothermal spherical isotropic galaxy has constant $V_{\rm rms}$ (\citealt{Binney_and_Tremaine(2008)}, denoted by $\sigma$ in their eq. 4.100), while in this work, the same correspondence between $\gamma_{\rm tot}=2$ and flat $V_{\rm rms}$ profile also works in our non-spherical case. By contrast, TNG galaxies show a gradually increasing but somehow shallower trend of $\gamma_{\rm rms}$ towards the high $\mathrm{B/T}$ and high $\gamma_{\rm tot}$ end.

\subsection{Relations at fixed $\sigma_{\rm e}$}
\label{sec:result3}
\begin{figure*}
\centering
\includegraphics[width=1.8\columnwidth]{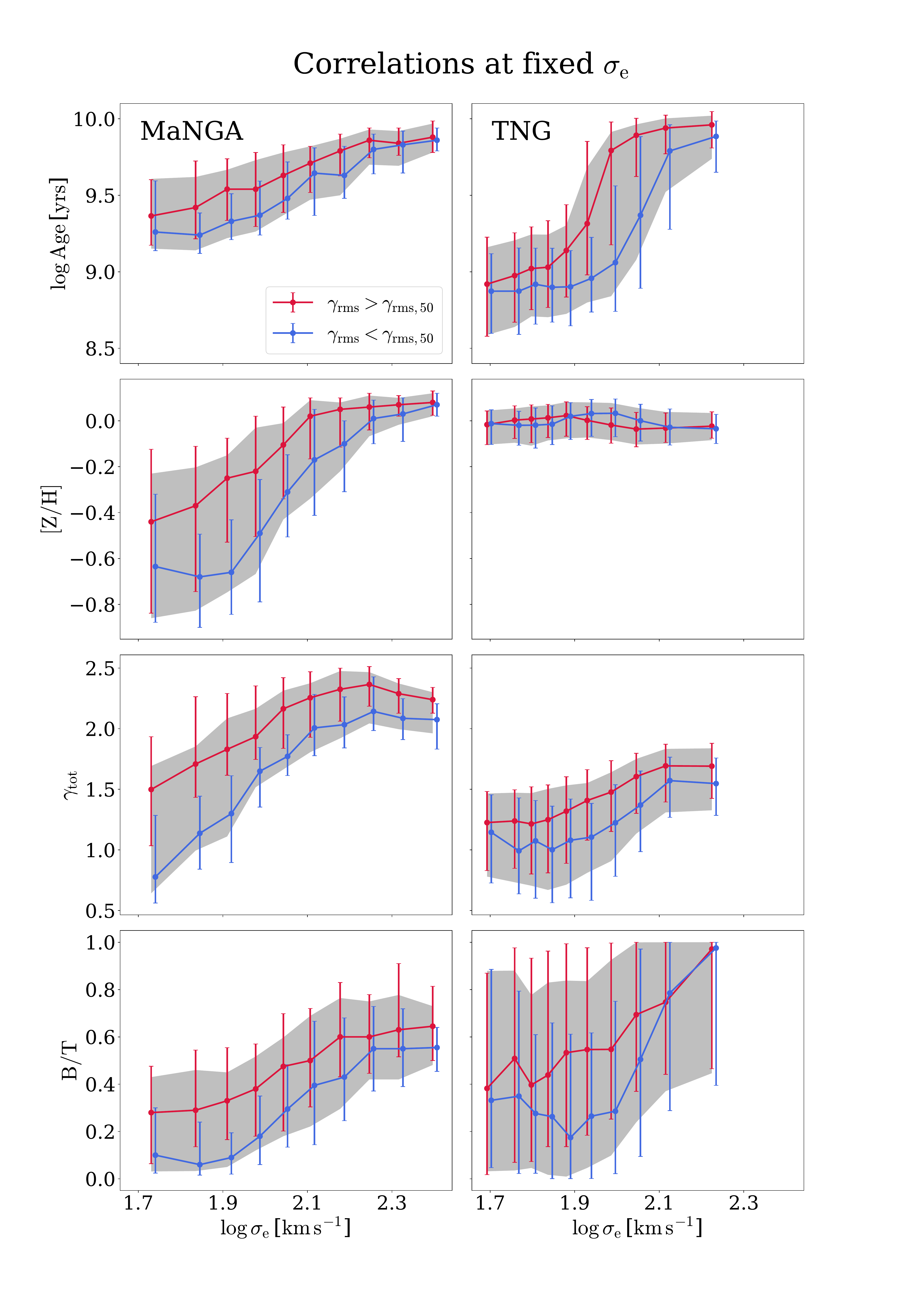}
\caption{The trend of galaxy properties as a function of $\log\sigma_{\rm e}$. Results of the stellar age, metallicity, the total density slope $\gamma_{\rm tot}$, and the bulge fraction $\mathrm{B/T}$ are shown from top to bottom. The left panels are for MaNGA galaxies and the right panels are for TNG galaxies. In each panel, galaxies in each $V_{\rm rms}$ bin are divided into two subgroups ($\gamma_{\rm rms}>\gamma_{\rm rms,50}$ and $\gamma_{\rm rms}<\gamma_{\rm rms,50}$, where $\gamma_{\rm rms,50}$ is the median value of $\gamma_{\rm rms}$ in each $V_{\rm rms}$ bin). Red and blue dots denote the median values of galaxy properties as shown in the Y-axis of the two subgroups ($\gamma_{\rm rms}>\gamma_{\rm rms,50}$ and $\gamma_{\rm rms}<\gamma_{\rm rms,50}$) with error bars indicating the range from the 16th to the 84th percentiles ($1\sigma$). The grey shaded region indicates the same $1\sigma$ range for the relations between galaxy properties and $\sigma_{\rm e}$ for all MaNGA (left panels) and TNG (right panels) galaxies.}
\label{fig:fixedsigma}
\end{figure*}

The results in Section~\ref{sec:result2} are generally consistent with the previous studies \citep{Cappellari_et_al.(2013b),Scott_et_al.(2017),Li_et_al.(2018)}, in which stellar age, metallicity, and the gradient of $V_{\rm rms}$ are found to have similar distributions on the mass-size plane, indicating a correlation among them. However, it is still unclear whether the relations are only driven by $\sigma_{\rm e}$, or in other words, whether the gradient of $V_{\rm rms}$ still correlates with galaxy properties at fixed $\sigma_{\rm e}$.

To answer this question, we first divide galaxies in our sample into 10 bins according to their $\log\,\sigma_{\rm e}$, with the same number of galaxies in each bin. Then galaxies in each bin are further divided into two subgroups according to their $V_{\rm rms}$ gradient $\gamma_{\rm rms}$ ($\gamma_{\rm rms}>\gamma_{\rm rms,50}$ and $\gamma_{\rm rms}<\gamma_{\rm rms,50}$, where $\gamma_{\rm rms,50}$ is the median value of $\gamma_{\rm rms}$ in each $V_{\rm rms}$ bin). Finally, we calculate the median values of galaxy properties for galaxies of both two subgroups in each bin. 

Fig.~\ref{fig:fixedsigma} shows the trends of galaxy properties as a function of $\log\sigma_{\rm e}$ for both subgroups ($\gamma_{\rm rms}>\gamma_{\rm rms,50}$ and $\gamma_{\rm rms}<\gamma_{\rm rms,50}$). It can be clearly seen that, although with large scatters, MaNGA galaxies with different $V_{\rm rms}$ gradients ($\gamma_{\rm rms}$) appear to show systematic differences between each other. At fixed $\sigma_{\rm e}$, galaxies with larger $\gamma_{\rm rms}$ (red lines) are typically older, more metal-rich, more bulge-dominated, and have steeper mass profiles. 

To quantitatively describe the discrepancy between the two subgroups ($\gamma_{\rm rms}>\gamma_{\rm rms,50}$ and $\gamma_{\rm rms}<\gamma_{\rm rms,50}$), we employ a statistical analysis to estimate the probability of the difference being statistically significant. Thus, we set our null hypothesis $H_0$ to be: the two subgroups ($\gamma_{\rm rms}>\gamma_{\rm rms,50}$ and $\gamma_{\rm rms}<\gamma_{\rm rms,50}$) do not have systematic difference in age (or other parameters). In the $i$-th $\log\,\sigma_{\rm e}$ bin, if the discrepancy we see for the two subgroups arises by chance, the age (or other investigated parameters) difference should follow a normal distribution with its mean value being 0 and its variance being the combination of the errors of the two subgroups:
\begin{equation}
    \sigma_{i}^2=\sigma_{i,\gamma_{\rm rms}>\gamma_{\mathrm{rms},50}}^2+\sigma_{i,\gamma_{\rm rms}<\gamma_{\mathrm{rms},50}}^2,
\end{equation}
where $\sigma_{i,\gamma_{\rm rms}>\gamma_{\mathrm{rms},50}}$ and $\sigma_{i,\gamma_{\rm rms}<\gamma_{\mathrm{rms},50}}$ are the standard deviations of age (or other parameters) of the two subgroups in $i$-th bin. We randomly sample from the 10 (the number of $\log\,\sigma_{\rm e}$ bins) normal distributions, and define:
\begin{equation}
    \Delta = \sum_{i=1}^{10}\,\frac{\delta_{i}}{\sigma_{i}},
\end{equation}
where $\delta_{i}$ is the value randomly sampled from the normal distribution of the $i$-th bin. Then we compare $\Delta$ with the observed relative difference ($\Delta_{\rm ob}\equiv \sum_{i=1}^{10}\,\delta_{\mathrm{ob},i}/\sigma_{i}$, where $\delta_{\mathrm{ob},i}$ is the difference we see in Fig.~\ref{fig:fixedsigma} of the two subgroups in the $i$-th bin). We want to evaluate the probability of seeing all sampled $\delta_{i}$’s being positive and $\Delta$ greater than the observed value, $\Delta_{\rm ob}$, by random fluctuations. 

\begin{figure}
\centering
\includegraphics[width=1\columnwidth]{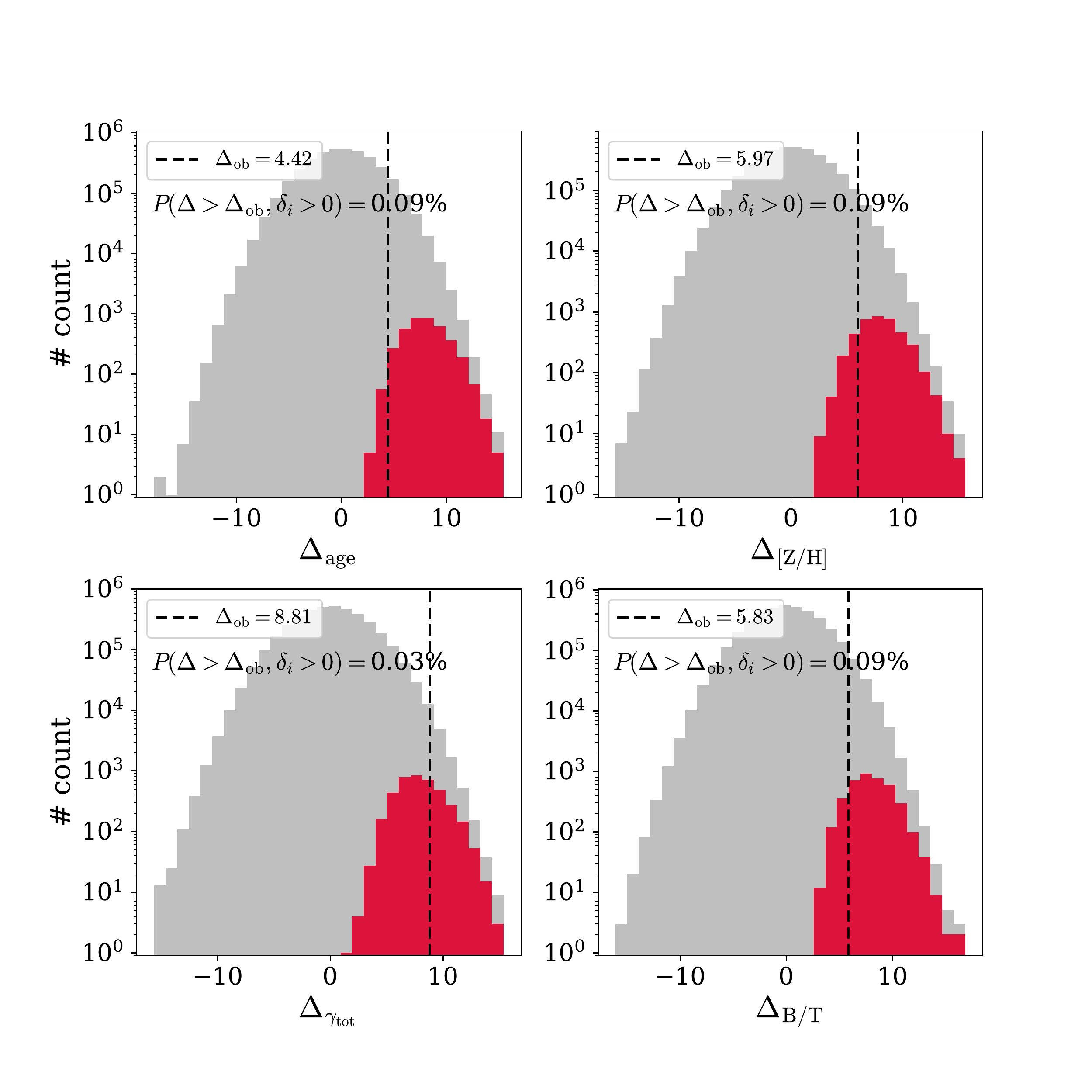}
\caption{The distributions of sampled values for age (upper left), metallicity (upper right), $\gamma_{\rm tot}$ (bottom left), and $\mathrm{B/T}$ (bottom right). In each panel, the grey histograms indicate the distribution of sampled values, while the red histograms indicate the distribution of sampled values which satisfy the criterion of all sampled $\delta_{i}>0$. The black dashed line indicates the $\Delta_{\rm ob}$ for age, metallicity, $\gamma_{\rm tot}$, and $\mathrm{B/T}$. The probability $P(\Delta>\Delta_{\rm ob},\,\delta_{i}>0)$ in each panel indicates the fraction of sampled values which satisfy both $\Delta>\Delta_{\rm ob}$ and all sampled $\delta_{i}>0$.}
\label{fig:statistical}
\end{figure}

To do this, we perform Monte Carlo samplings mentioned above by $4\times 10^6$ times, and find that $P(\Delta>\Delta_{\rm ob},\,\delta_{i}>0) \sim 0.09\%$ for galaxy age, metallicity and $\mathrm{B/T}$, and $\sim0.03\%$ for $\gamma_{\rm tot}$ (see Fig.~\ref{fig:statistical} for the distributions of sampled values). Given such low probabilities of our observed differences arising by chance, we reject our null hypotheses and conclude that there is indeed systematic difference in both stellar population properties and structural parameters of galaxies with different $\gamma_{\rm rms}$ at fixed $\sigma_{\rm e}$.

In comparison, the two subgroups of TNG galaxies also show discrepancies in the mass density slope, the bulge fraction, and the stellar age with $P(\Delta>\Delta_{\rm ob},\,\delta_{i}>0)$ being $\sim 0.1\%$, $\sim 0.1\%$, and $\sim 0.09\%$, respectively. For metallicity, however, the difference between the two subgroups is again completely absent in the simulations. In particular, TNG galaxies span a rather limited range of $\mathrm{[Z/H]}$ compared to the observations, making any comparison with the observations flawed from the start.

The results seen from Fig.~\ref{fig:fixedsigma} indicate that the relations between $\sigma_{\rm e}$ and galaxy properties are not only driven by $\sigma_{\rm e}$, but also by the gradient of $V_{\rm rms}$. The fact that the gradient of $V_{\rm rms}$ contains information of both stellar populations and structural properties of galaxies, which are not fully accounted for by $\sigma_{\rm e}$, has not been addressed before.

\section{Conclusion and Discussion}
\label{sec:conclusion}
In this work, we study the relations between the gradient of $V_{\rm rms}$ ($\equiv \sqrt{V^2+\sigma^2}$) and galaxies properties (i.e. the mass-weighted total density slope $\gamma_{\rm tot}$, the bulge fraction $\mathrm{B/T}$, galaxy age $\log\mathrm{Age}$, and metallicity $\mathrm{[Z/H]}$) with 1339 galaxies from the MaNGA sample released by SDSS DR14, and make a comparison with the hydrodynamical simulations, the IllustrisTNG (TNG) simulations. 

We employ $\gamma_{\rm rms}$ ($\equiv \sigma(R_{\rm e}/4)/\sigma_{\rm e}$) to represent the gradient of $V_{\rm rms}$ and find that $\gamma_{\rm rms}$ changes systematically on the mass-size plane roughly along the direction of $\sigma_{\rm e}$, consistent with C13b. Besides, $\gamma_{\rm rms}$ increases with $\log\sigma_{\rm e}$ below $\log \sigma_{\rm e}\approx 2.1$. At high $\sigma_{\rm e}$ end, the $V_{\rm rms}$ gradient keeps nearly unchanged at $\gamma_{\rm rms}\approx1.05$, indicating a saturation in the $V_{\rm rms}$ slope for bulge-dominated galaxies ($\mathrm{B/T}\gtrsim 0.5$). Most galaxies ($\sim 86\%$) in TNG appear to have $\gamma_{\rm rms}<1$, which means a decreasing trend of $V_{\rm rms}$ from inner to outer regions in galaxies, relative to $44\%$ in MaNGA. It may be due to the overly-strong AGN feedback in TNG \citep{Wang_Y._et_al.(2019),Wang_Y._et_al.(2020),Lu_et_al.(2020)}. 

We find that $V_{\rm rms}$ gradients $\gamma_{\rm rms}$ are closely related (in a non-linear way) to both the total mass density gradients $\gamma_{\rm tot}$ and, more weakly, to the bulge fraction $\mathrm{B/T}$, which is a more uncertain quantity. We confirm the clear trends of $\gamma_{\rm rms}$ with $\sigma_{\rm e}$, age and metallicity (e.g. fig. 22 in the review by \citealt{Cappellari(2016)}). The correlation of $\gamma_{\rm rms}$ with age is more shallow in TNG than in the observations and our clear empirical correlation of $\gamma_{\rm rms}$ with [Z/H] is completely absent in TNG.

The main goal of this paper is to look for trends in galaxy properties at fixed $\sigma_{\rm e}$, which were impossible to investigate from smaller samples than MaNGA. We found that at fixed $\sigma_{\rm e}$ the stellar population clearly depends on the density gradients as here quantified by $\gamma_{\rm rms}$. In particular, at fixed $\sigma_{\rm e}$, galaxies with larger $\gamma_{\rm rms}$ are still older and more metal-rich. It means that both stellar populations and structural properties of galaxies are not fully accounted for by $\sigma_{\rm e}$ and the gradient of $V_{\rm rms}$ also contains these information. In TNG, galaxy age, $\gamma_{\rm tot}$, and $\mathrm{B/T}$ show qualitatively similar trends as we derive from the MaNGA data, but the trends at fixed $\sigma_{\rm e}$ of $\gamma_{\rm rms}$ with $\mathrm{[Z/H]}$, which are clearly visible in MaNGA, are again completely absent in the simulations. It implies that the relation between metallicity and stellar age (and also other galaxy properties) seen in observations is not met in TNG.

This new empirical evidence that galaxy properties still correlate with the $V_{\rm rms}$ gradient $\gamma_{\rm rms}$ at fixed $\sigma_{\rm e}$ provides an extra constraint on our understanding of galaxy quenching. In the current state-of-the-art simulations (e.g. the IllustrisTNG simulations), the relations between galaxy properties and $\gamma_{\rm rms}$ are not well recovered, which shows the power of investigating this extra quantity $-$ it can be used to validate the simulations in terms of the relations between stellar population and dynamical properties of galaxies.

\section*{Acknowledgements}
This work is partly supported by the National Key Basic Research and Development Program of China (No. 2018YFA0404501 to SM), by the National Natural Science Foundation of China (Grant Nos. 11821303, 11761131004 and 11761141012 to SM, 11903046 to JG), and by the Beijing Municipal Natural Science Foundation (No. 1204038 to JG). RL is supported by National Natural Science Foundation of China (Nos. 11773032 and 118513) and the NAOC Nebula Talents Program.

%%%%%%%%%%%%%%%%%%%%%%%%%%%%%%%%%%%%%%%%%%%%%%%%%%

\bibliographystyle{mnras}
\bibliography{ref} 

\label{lastpage}
\end{document}